\documentclass[referee]{aa}
\usepackage[varg]{txfonts}    
\usepackage{natbib}
\usepackage{url}             
\usepackage{graphicx}
\usepackage{float}
\newcommand{\beq}{\begin{equation}}
\newcommand{\eeq}{\end{equation}}
\newcommand{\beqa}{\begin{eqnarray}}
\newcommand{\eeqa}{\end{eqnarray}}
\newcommand{\beqann}{\begin{eqnarray*}}
\newcommand{\eeqann}{\end{eqnarray*}}
\bibpunct{(}{)}{;}{a}{}{,} 

\usepackage{textcomp}
\usepackage{wrapfig}

\begin{document}

\titlerunning{Rotating Network Jets}
\title{Rotating Network Jets \textbf{in the quiet Sun} as Observed by IRIS}

\offprints{P.~Kayshap}

\author{P.~Kayshap\inst{\ref{inst1}}
\and K.~Murawski\inst{\ref{inst1}}
\and A.K.~Srivastava\inst{\ref{inst2}}
\and B.N. Dwivedi\inst{\ref{inst2}}
}

\offprints{P.~Kayshap}

\institute{Group of Astrophysics, University of Maria Curie-Sk{\l}odowska, ul. Radziszewskiego 10, 20-031 Lublin, Poland \label{inst1}
\and
Department of Physics, Indian Institute of Technology (Banaras Hindu University), Varanasi-22105, India \label{inst2}
}

\abstract
{}
{
We perform a detailed observational analysis of network jets to understand their kinematics, rotational motion and underlying triggering mechanism(s). \textbf{We have analyzed the quiet-Sun (QS) data.}
}
{
IRIS high resolution imaging and spectral observations (SJI: Si~{\sc iv} 1400.0~{\AA}, Raster: Si~{\sc iv} 1393.75~{\AA}) are used to analyze the omnipresent rotating network jets in the transition-region (TR). In addition, we have also used Atmospheric Imaging Assembly (AIA) onboard Solar Dynamic Observation (SDO) observations.}
{The statistical analysis of fifty-one network jets is performed to understand various their mean properties, e.g., apparent speed (140.16$\pm$39.41 km s$^{-1}$), length (3.16$\pm$1.18 Mm), lifetimes (105.49$\pm$51.75 s). The Si~{\sc iv} 1393.75~{\AA} line has secondary component along with its main Gaussian, which is formed due to the high-speed plasma flows (i.e., network jets). The variation of Doppler velocity across these jets (i.e., blue shift on one edge and red shift on the other) signify the presence of inherited rotational motion. The statistical analysis predicts that the mean rotational velocity (i.e., $\Delta$V) is 49.56 km s$^{-1}$. The network jets have high angular velocity in comparison to the other class of solar jets. 
}
{The signature of network jets are inherited in TR spectral lines in terms of the secondary component of the Si~{\sc iv} 1393.75~{\AA} line. The rotational motion of network jets is omnipresent, which is reported firstly for this class of jet-like features. The magnetic reconnection seems to be the most favorable mechanism for the formation of these network jets.}

\keywords{Sun: activity - Sun: corona - Sun: transition region}

\titlerunning{Rotating Motion of Network Jets}
\authorrunning{P.~Kayshap et al.}

\maketitle

\section{Introduction}
Various kinds of jet-like structures (i.e., spicules, chromospheric anemone jets, macrospicules, surges, X-ray/UV/EUV jets, etc.) are episodically present in the solar atmosphere. Study of these jet-like structures is one of the important areas in solar physics research using observations/numerical simulations. Therefore, our depth of knowledge about them (e.g., formation, evolution, plasma properties, etc.) is continuously improving (e.g., \citealt{Wilhelm2000,DePon2004,DePon2007,Murawski2010,Per2014, Shibata2007,Nisi2008,Bohlin1975, Wilhelm2000, Kamio2007,Murawski2011,Kayshap2013, Yokoyama1995,Sch1995, Yokoyama1996,Canfield1996,Chae1999,Cirtain2007,Abhi2011,Mortan2012,Kayshap2013a,Mark2015,Mulay2016a,Mulay2016b,Rao17}). \textbf{It is to be noted that the jet acceleration mechanisms strongly depend on the height of the drivers (e.g. \citealt{Shibata2007,Takasao2013})}. Recently, \cite{Rao2016} have reviewed various aspects of the solar coronal jets in observations, theory and numerical modeling.\\
Recently, one more class of jet-like structures is discovered using IRIS coronal hole (CH) observations (network jets; \citealt{Tian2014}). The network jets are typically the TR phenomena. The TR, which is an interface between relatively cool chromosphere ($\sim$ 6$\times$10$^{3}$ K) and hot corona ($\sim$ 10$^{6}$ K), is a very complex/dynamic layer (\citealt{Kay15}). TR is not studied with the fine details due to the unavailability of its high-resolution observations. Now, the IRIS mission (i.e., slit-jaw-images (SJIs) $\&$ spectral observations) is particularly dedicated to TR (\citealt{DePon2014}). The networks jets are well observed in the TR filters (e.g., IRIS/SJI: C~{\sc ii} 1330~{\AA} and Si~{\sc iv} 1400~{\AA}) as they are the most dominant/prominent features of the TR. The network jets have the apparent speed of 80-250 km s$^{-1}$ with lifetimes of 20-80 s and length of 4-10 Mm (\citealt{Tian2014}). The width of these network jets is $\leq$300 km as reported by \cite{Tian2014}. In addition to the properties of these network jets, the magnetic reconnection as inferred from very high speed and associated footpoint brightenings, is reported as a triggering mechanism for these jets (\citealt{Tian2014}). In an another work, it is reported that network jets also occur in the quiet-Sun (QS), and certainly not only in the coronal holes (CHs) (\citealt{Narang2016}). Interestingly, they have also reported that the network jets are faster and longer in CH than the QS. This can be directly attributed to a difference in the magnetic field configuration between QS and CH regions, as well as the height of the TR.\\
The rotating motion is a very well known property of the jet-like structures. The specific Doppler shifts pattern of jet-like structures (i.e., blue on one edge and red on the other edge) predicts the presence of rotating motion (e.g., \citealt{Pike1998,Curdt2011,Mark2015}). The magnetic reconnection between emerging magnetic bi-pole and pre-existing magnetic field produces the rotating jets (e.g., \citealt{Fang2014,Mark2015,Lee2015}). In addition, the photospheric horizontal motions can also add the twist to the magnetic fields which finally produce the helical jet through the magnetic reconnection (e.g., \citealt{Par2009,Pariat2010}). Recently, \citealt{DePon2014} have reported the prevalence of small-scale twist in the solar chromosphere and TR, which is very important from the perspective of the heating of lower atmosphere. \\
In the present work, we use IRIS spectroscopic/imaging observations for the statistical analysis of network jets in the context of rotating motion, and other associated properties (e.g., speed, length, life-time). The observations and data analysis are presented in the Sect~\ref{section:obs}. The section~\ref{section:results} describes observational results. Discussion and conclusions are described in the last section. 
\section{Observations and Data-Analysis} \label{section:obs}
IRIS mission provides the high-resolution imaging/spectroscopic observations from the photosphere up to the corona (\citealt{DePon2014}). The C~{\sc ii} 1330~{\AA} and Si~{\sc iv} 1400~{\AA} imaging filters (slit-jaw images (SJI)) capture the emissions from transition-region (TR). The network jets are best seen in the TR, therefore, imaging observations from these two filters (i.e., C~{\sc ii} $\&$ Si~{\sc iv}) capture the dynamics of recently discovered network jets (\citealt{Tian2014}). We have used three different observations for the study of the dynamics of network jets. The details of the used observations are given in Table~\ref{table}. In this table, we have outlined necessary information (i.e., date $\&$ time, field-of-view (FOV) and exposure times of SJI and raster) about the used observations.\\
\begin{table*}[ht]
\centering
 \caption{The table shows the date $\&$ time, field of view (FOV) and exposure time (in seconds) for SJI and raster for all three sets of the observations. \label{table_obs}} 
\begin{tabular}{|c|c|c|c|}
\hline
Observation & Date-Time & FOV & Exposure Time(SJI/Raster; Seconds) \\               
\hline
Obs$\_$A & 14.12.2014 (15:38-16:35) & 134$"$$\times$119$"$ & 20.0/4.0 \\
\hline
Obs$\_$B & 24.09.2014 (18:09-20:17) & 119$"$$\times$119$"$ & 38.0/8.0 \\
\hline
Obs$\_$C & 23.09.2014 (07:59-10:56) & 60$"$$\times$65$"$   & 11.0/4.0 \\
\hline
\end{tabular}
\label{table}
\end{table*}
The Si~{\sc iv} as well as C~{\sc ii} spectral lines originate from the TR. The C~{\sc ii} resonance lines (i.e., 1334.53~{\AA} and 1335.71~{\AA}) are optically thick lines, however, the Si~{\sc iv} 1393.75~{\AA} is an optically thin line under the normal conditions. Therefore, we have used Si~{\sc iv} 1393.75~{\AA} line to infer the Doppler velocities of these network jets. The level-2 data files from IRIS are the standard scientific products (\citealt{DePon2014}), which are used in the present analysis. The imaging and spectrogram are already aligned with each other in these level-2 files, however, we have also checked this alignment using fiducial marks and found that data are well aligned. In addition, we have also used the observations from AIA (\citealt{Lemen2012}). IRIS/SJI C~{\sc ii} 1330~{\AA} filter captures the significant emissions from the continuum. Therefore, we have used the cross correlation between IRIS/SJI C~{\sc ii} 1330~{\AA} and AIA 1600~{\AA} filter observations for the alignment between IRIS and AIA observations. The IRIS/SJI Si~{\sc iv} 1400~{\AA} observations are used to draw the various properties (e.g., speed, life-time and length) of these network jets. We utilize the space-time technique to draw various properties of these network jets.\\
The single Gaussian can easily characterize the Si~{\sc iv} 1393.75~{\AA} due to its optically thin nature. However, almost all of the Si~{\sc iv} 1393.75~{\AA} spectral profiles are double peak or asymmetric in the vicinity of network jets. So, these profiles are not well fitted by Single Gaussian. The double Gaussian provides a more reliable fitting on the observed spectral profiles. The estimation of the rest wavelength is another crucial issue as it can directly affect the estimated Doppler velocity. To estimate the rest wavelength, very quiet-area is selected from each raster. The wavelength from the averaged spectra of quiet-areas, which are free from any kind of dynamics, represents the rest wavelength of Si~{\sc iv} 1393.75~{\AA}.
\section{Observational Results}\label{section:results}
IRIS high-resolution spectral/imaging observations are important to diagnose very dynamic network jets of TR. We have visually identified total 51 network jets (19-Obs$\_$A; 19-Obs$\_$B; 12-Obs$\_$C) from all three different observations. The selection of all these jets is made in such a way that each jet should be very well isolated from the others. We have also adopted that the each jet should be visible in three image frames. We have adopted this criterion to avoid any possible error associated with the jet identification in our analysis. Under the light of these criteria, we have selected well isolated 51 network jets for this work. As stated earlier that the IRIS/SJI Si~{\sc iv} 1400~{\AA} $\&$ Si~{\sc iv} 1393.75~{\AA} spectral line are used to diagnose the Doppler velocity and other associated properties ( e.g., life-time, speed and length) of these network jets.  
\subsection{Temporal Evolution and Kinematics of the Network Jets}
\subsubsection{Temporal Evolution of Network Jets}\label{sect:evolution}
\begin{figure*}
\centering
\mbox{
\includegraphics[trim = 1.5cm 2.0cm 4.5cm 3.5cm, scale=1.1]{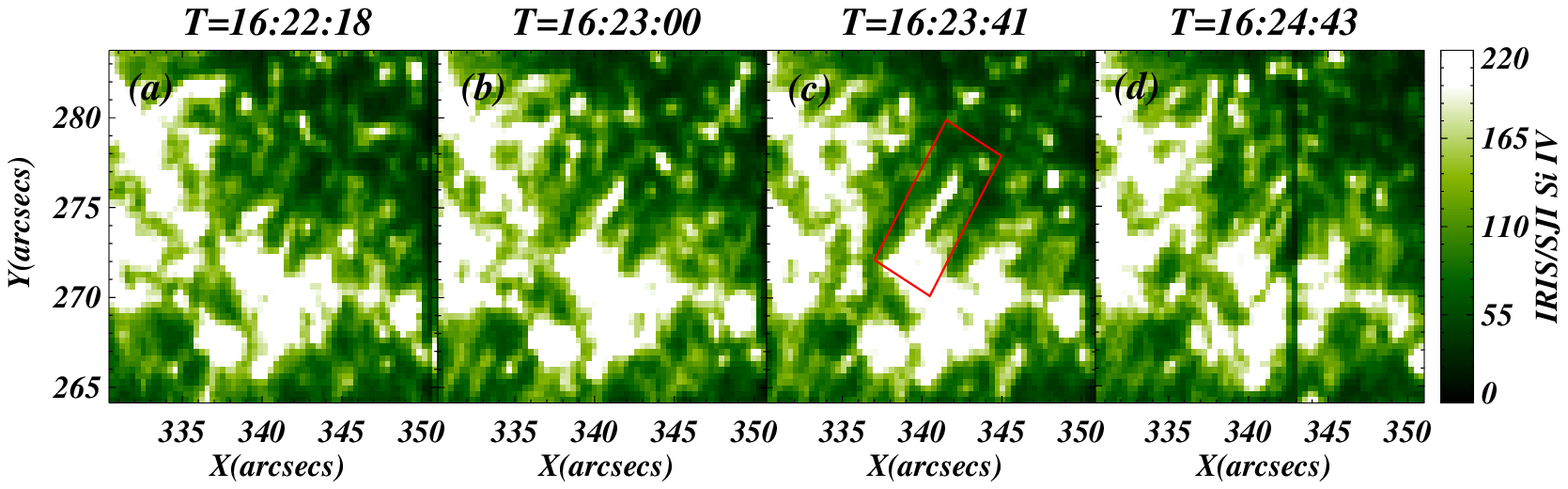}
}
\mbox{
\includegraphics[trim = 1.5cm 2.0cm 4.5cm 5.5cm, scale=1.1]{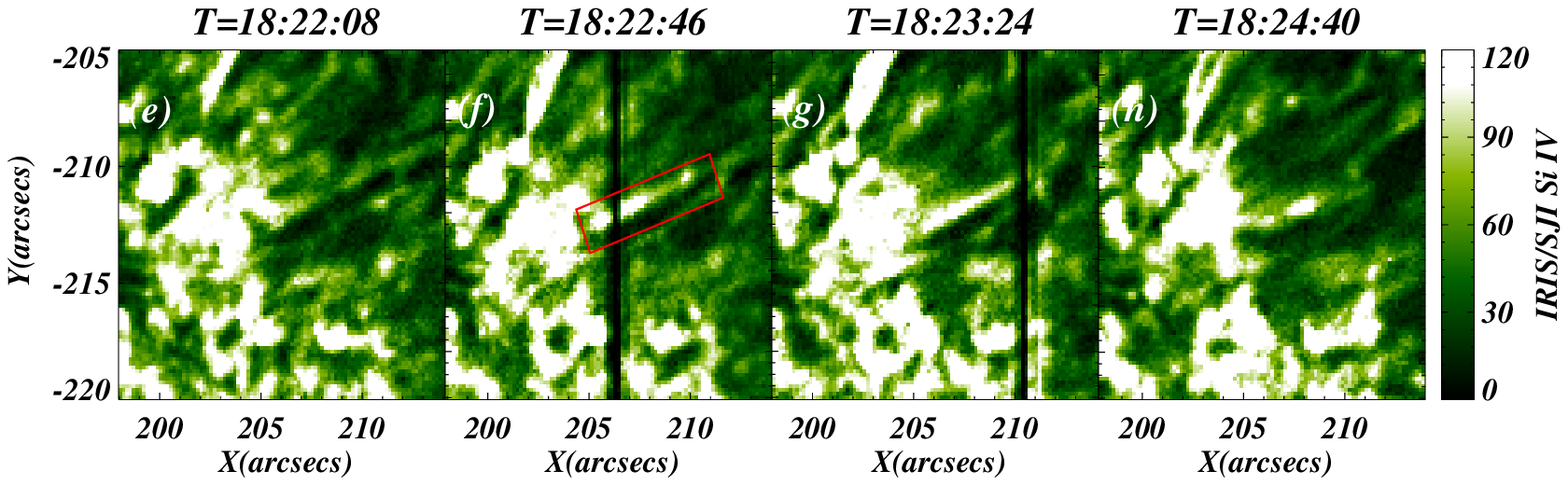}
}
\mbox{
\includegraphics[trim = 1.5cm 3.0cm 4.5cm 5.5cm, scale=1.1]{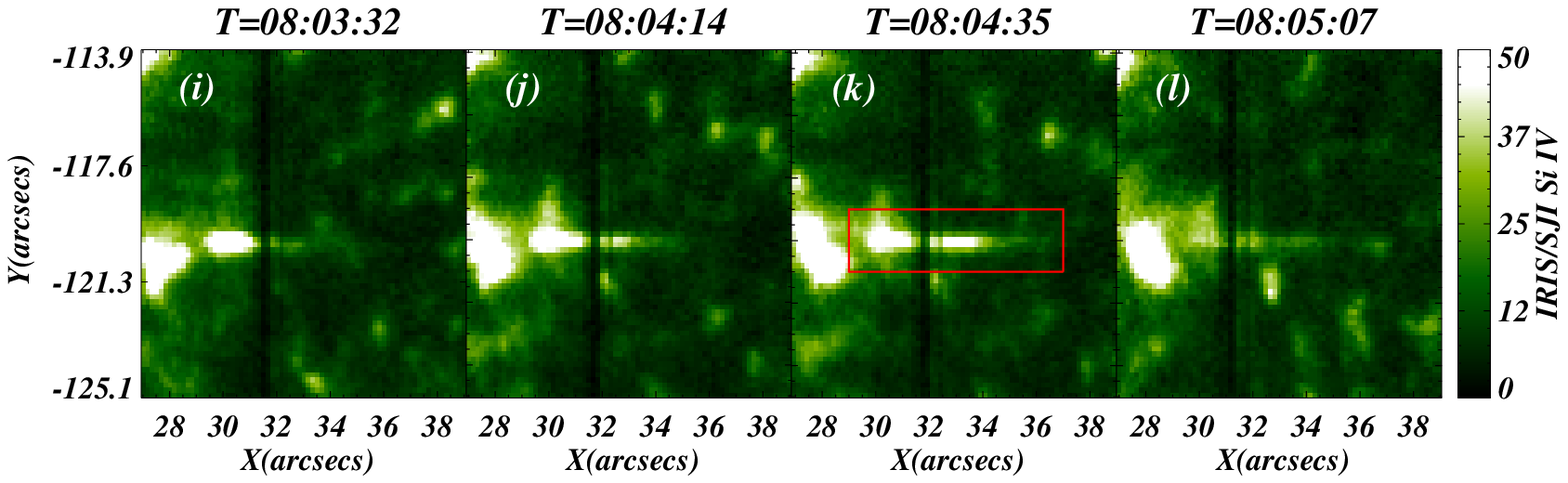}
}
\caption{\small The top-row shows the evolution of a network jet (panels a-d) taken from first observation (Obs$\_$A). The jet originates at 16:22:18~UT on 14 December 2014 from the edge of the magnetic network (panel a) with its maximum phase around 16:23:41 UT (panel c; as outlined by red rectangular box). The jet fades from the view in the last phase (panel d). The middle row shows the evolution of another network jet taken from second observation (Obs$\_$B) and the jet is shown by red rectangular box. Similarly, the bottom row depicts the evolution of one more network jet which is taken from third observation.}
\label{fig:jet_evol}
\end{figure*}

In Fig.~\ref{fig:jet_evol}, we have shown the evolution of three different network jets in IRIS/SJI Si~{\sc iv} 1400~{\AA}. 
The top row (i.e., panels a to d) of Fig.~\ref{fig:jet_evol} shows evolution of a network jet that is taken from first observation (i.e., Obs$\_$A; table~\ref{table}). The network jets originate from the bright patches (magnetic network; \citealt{Tian2014,Narang2016}). The extended bright patch is visible in the vicinity of this network jet in IRIS/SJI Si~{\sc iv} 1400~{\AA} (cf., panels a to d). The jet started around 16:12:28~UT from one 
edge of bright patch (panel a). Further, the jet is growing and acquires the maximum phase around 16:23:41~UT (panel c). The red rectangular box outlines the network jet. At later times, the jet fades from the view as visible in the last panel of top row. The network jet reaches up to 2.8 Mm from its base within the life-time of approximately 244.0 seconds.\\
In the middle row, we have shown the evolution of another network jet (panels e to h) that is taken from second observation (i.e., Obs$\_$B; table~\ref{table}). The bright patch is visible in the vicinity of this network jets, which is similar to the first jet. The jet starts to form around 18:22:08~UT on 24 January 2014 from one edge of the magnetic network. The jet further grows obliquely from its initiation site, which is outlined by the red rectangular box (panel f). The black vertical line shows the slit position, which is used to take the spectra. The jet attains its maximum phase after around 136 seconds (18:23:24; panel g). In the later times, the network jet fades from the view (panel h). The network jet reaches up to 5.7 Mm within its total life time of 197.0 seconds.\\ 
Finally, in the bottom row, we show the temporal evolution of one more network jet (panels i to l) which is taken from third observation (i.e., Obs$\_$C; table~\ref{table}). The bright patch is visible in the vicinity of jet's base, which justify the base of jet is located in the magnetic network. The jet starts around 08:03:32~UT from the brightened area (panel i). The jet is evolving further nearly in the vertical direction. The red rectangular box outlines the jet and black vertical line is the slit position (panel k). The maximum phase of this particular jet occurs around 08:04:35~UT (panel k). The jet attains its maximum length of 3.7 Mm within its total lifetime of around 95 seconds. The jet fades from view in the decay phase (panel l). So, we can say that network jets follow the typical evolution and fades from the view in the decay phase. Most importantly, these network jets always have brightened base and originate from the magnetic networks. 
\begin{figure*}
\centering
\mbox{
\includegraphics[trim = 2.0cm 1.5cm 4.5cm 1.5cm, scale=1.0]{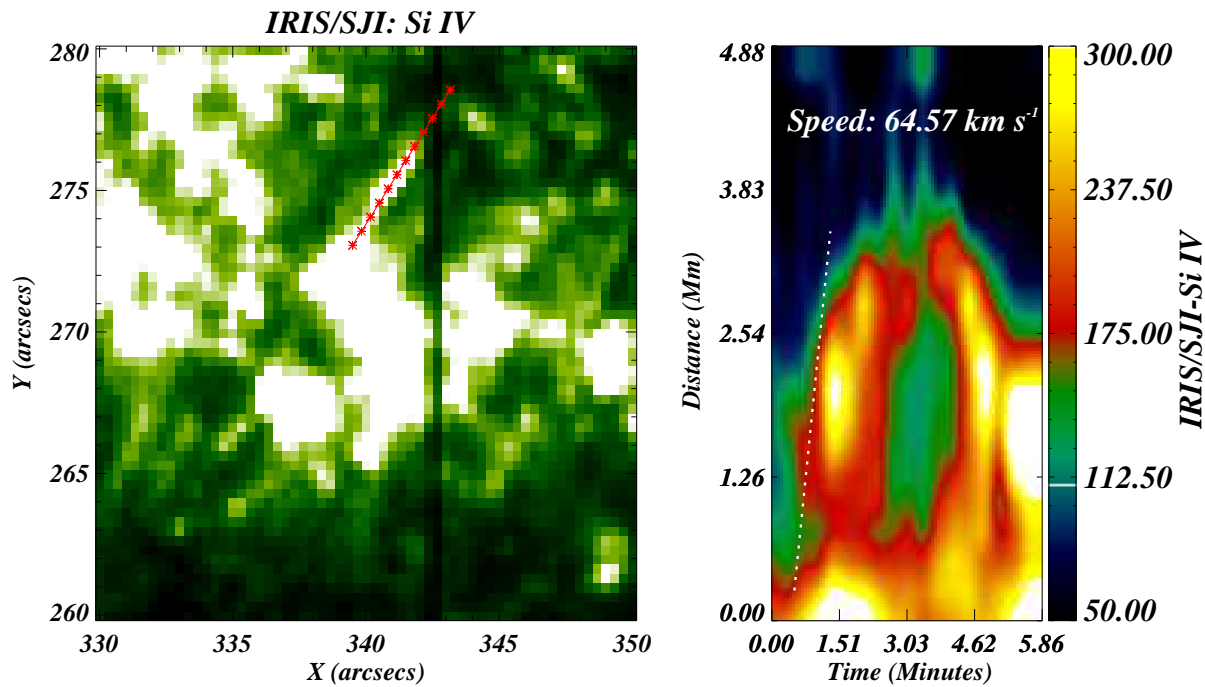}
}
\caption{\small Left-panel: SJI of Si~{\sc iv} 1400~{\AA} pass band of IRIS during the maximum phase of the jet along the selected path (over plotted red plus sign), which is used to produce the height-time diagram. Right-panel: height-time diagram of the network jet, which shows that the speed and life-time of the network jet are $\sim$ 64.57 km s$^{-1}$ and 244.0 seconds, respectively.}
\label{fig:ht_map}
\end{figure*}
\subsubsection{Kinematics of Network Jets}
The space-time technique is used to evaluate the kinematics of these network jets (e.g., speed, height and life-time). The Fig.~\ref{fig:ht_map} shows the height-time diagram for the jet from Obs$\_$A (first row; figure~\ref{fig:jet_evol}). The slit position is over-plotted on IRIS/SJI 1400~{\AA} intensity image by red asterisk signs, which is shown in the left-panel of Fig.~\ref{fig:ht_map}. Using this path along the network jet, we have used 3 pixels in the transverse direction to create the space-time diagram of this network jet (right-panel of Fig.~\ref{fig:ht_map}). We have drawn a path (white-dashed line) on the space-time diagram to measure the speed of this jet, which is about 64.57 km s$^{-1}$. The life-time of the jet is 244 seconds, however, the jet reaches up to 4 Mm. The another network jet also appears from the same site, which indicates the concurrent energy release at the origin site.\\
\begin{figure*}
\centering
\mbox{
\includegraphics[trim = 2.0cm 0.0cm 4.5cm 0.0cm, scale=1.0]{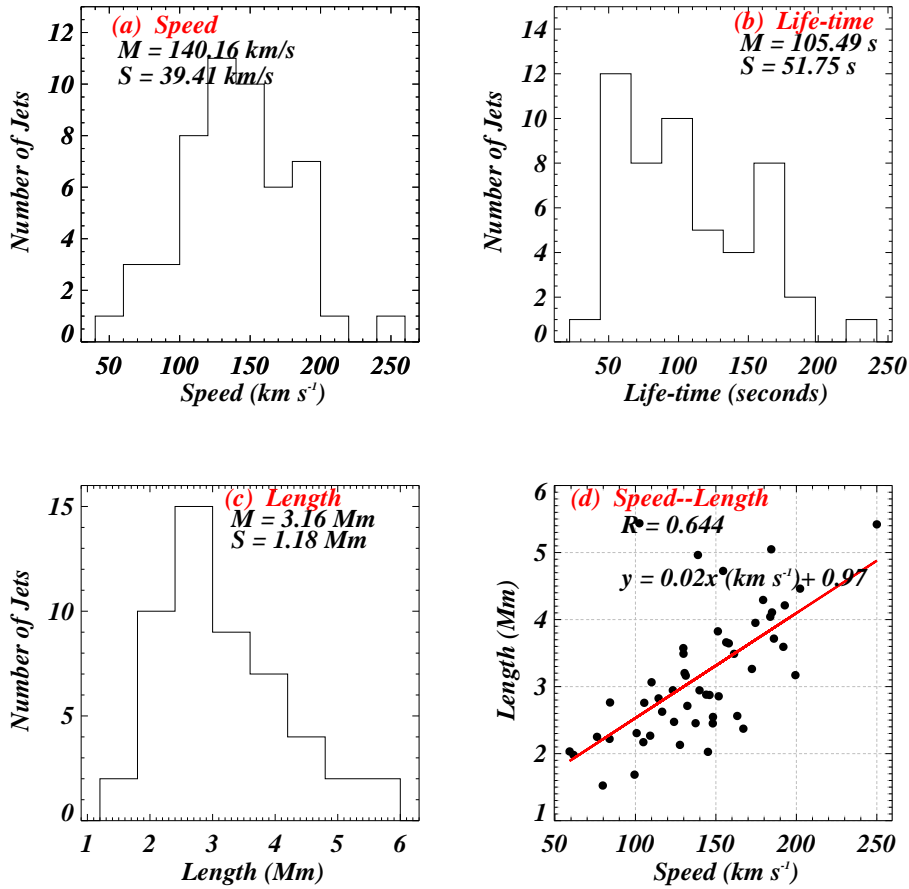}
}
\caption{\small The figure shows the distribution of apparent speed (panel a), life-time (panel b) and length (panel c) of the observed network jets. The mean speed of network jets is 140.16 km s$^{-1}$ with the mean lifetime of 105.49 seconds. The mean length is 3.16 Mm. Panel d shows the correlation between speed and length, which is positively correlated \textbf{as revealed by high pearson coefficient (R=0.644)}. }
\label{fig:hist_kin}
\end{figure*}
The parameters (e.g., life-time, speed and height) from all 51 network jets are estimated using space-time technique. We have produced the histogram 
for apparent speed (panel a; Fig.~\ref{fig:hist_kin}), lifetime (panel b; Fig.~\ref{fig:hist_kin}) and length (panel c; Fig.~\ref{fig:hist_kin}). The mean speed is 140.16 km s$^{-1}$ with its standard deviation of 39.41 km s$^{-1}$. However, the apparent speed can vary from 50.0 up to 200 km s$^{-1}$. The network jet can have life-time from 40.0 up to 250.0 seconds. However, the lifetime histogram of network jets predicts that mean life-time is 105.49 seconds with the standard deviation of 51.75 seconds. The histogram for the length of these network jets predicts the mean value of 3.16 Mm with the standard deviation of 1.18 Mm. It should also be noted that the length of these network jets can vary from 1.2 up to 5.8 Mm. In addition, we have also investigated the correlation between speed and length of these jets, which is shown in panel (d). It is clearly visible that speed and length of these jets are very well positively correlated. Such type of correlation is also investigated by \cite{Narang2016} and they have also reported the positive correlation between them. 
\subsubsection{Hot Counterparts of Network Jets}
Using IRIS/SJI Si~{\sc iv} 1400~{\AA} observations, Fig.~\ref{fig:jet_evol} shows temporal evolution of three network jets (cf., 
section~\ref{sect:evolution}). The various filters from AIA observations are used to investigate the possible hot counterparts of these network jets. However, the previous works are already reported that the network jets are strictly TR phenomena and their hot counterparts are not present in the solar atmosphere (e.g., \citealt{Tian2014,Narang2016}). 
\begin{figure*}
\centering
\mbox{
\includegraphics[trim = 1.5cm 3.0cm 4.5cm 5.5cm, scale=1.1]{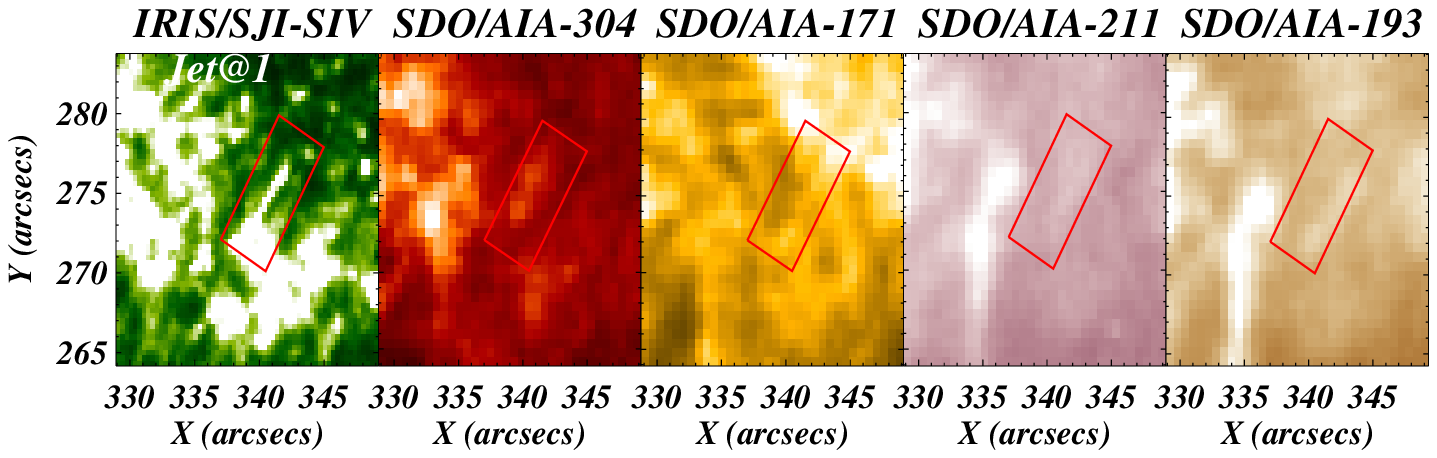}
}
\mbox{
\includegraphics[trim = 1.5cm 3.0cm 4.5cm 5.5cm, scale=1.1]{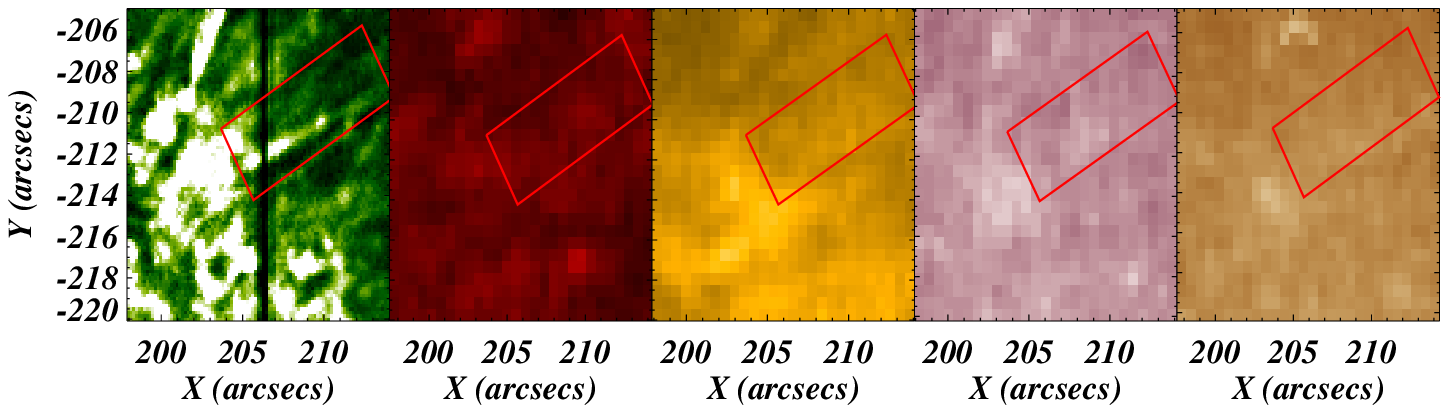}
}
\mbox{
\includegraphics[trim = 1.5cm 3.0cm 4.5cm 5.5cm, scale=1.1]{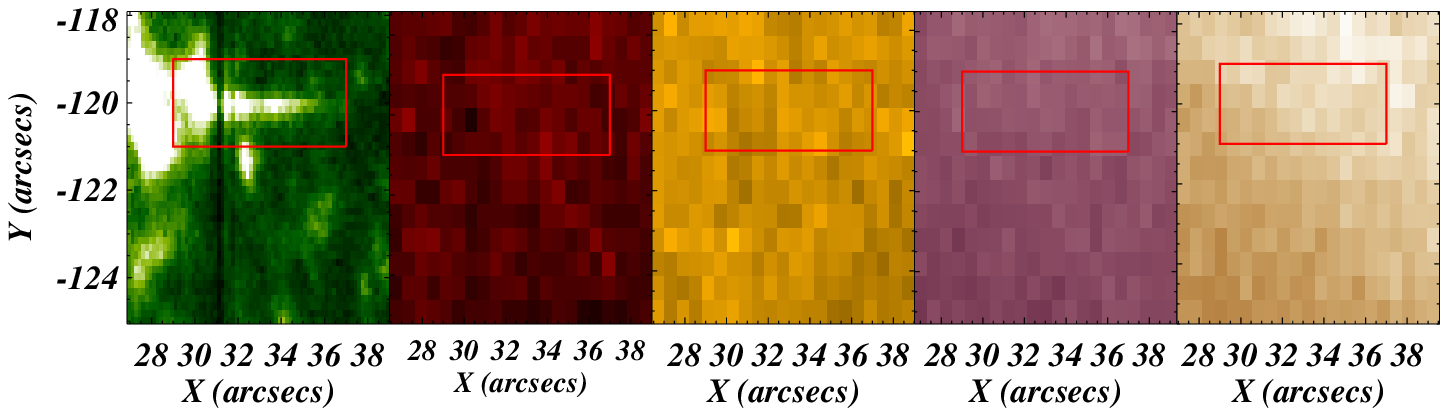}
}
\caption{\small Top row shows the network jet from Obs$\_$A (first jet; Fig.~\ref{fig:jet_evol}) in IRIS/SJI Si~{\sc iv} 1400~{\AA}, AIA-304~{\AA}, AIA 171~{\AA}, AIA 211~{\AA} and AIA 193~{\AA} respectively. The middle and bottom row the network jet from Obs$\_$B (second jet; Fig.~\ref{fig:jet_evol}) and from Obs$\_$C (third jet; Fig.~\ref{fig:jet_evol}) in the same IRIS and AIA filter. It is clearly evident that network jets do not possess hot counterparts.}
\label{fig:jet_aia}
\end{figure*}
We have investigated AIA 304~{\AA} (log T/K = ), AIA-171~{\AA} (log T/K = ), AIA-211~{\AA} (log T/K = ) and AIA-193~{\AA} (log T/K = ) to check the possibility of hot counterparts of the observed network jets. The top row of Fig.~\ref{fig:jet_aia} shows the IRIS/SJI Si~{\sc iv} 1400~{\AA}, AIA 304~{\AA}, 171~{\AA}, 211~{\AA} and 193~{\AA} images for a network jet taken from Obs$\_$A. It is clearly visible that jet is not observed in AIA filters (see red rectangular area). However, some traces of the network jet is visible in AIA~304~{\AA}. The temperature response of AIA~304~{\AA} filter is very wide and it can sample some low temperature plasma also. The middle row shows the second jet (Obs$\_$b) in IRIS and various AIA filters, which also predicts no hot counterparts of this network jet. Similarly, the bottom row shows the network jet in different filters (i.e., AIA and IRIS) and no hot counterparts for this network jet is visible too. Therefore, we can say that the network jets are clearly visible in the IRIS-SJI Si~{\sc iv} 1400~{\AA} (first panel in each row; Fig.~\ref{fig:jet_aia}), however, we do not see any signature of these network jets in the hot temperature filters. We have investigated the hot temperature filters for all network jets (i.e., 51 network jets) and we do not see the signature of network jets in these AIA filters. Therefore, these observations predict that network jets are typically the cool TR features, and our observations are consistent with the previously reported results on this aspect (\citealt{Tian2014,Narang2016}). 
\subsection{Rotational Nature of the Network Jets}
The rotational motion is an important property of the jet-like structures in the solar atmosphere. On the basis of Doppler velocity/Dopplergram analysis, it is reported that the typical coronal/chromospheric jets reveal blue-shifts at its one edge while the other side plasma experience red-shifts \cite[e.g.,][and references cited therein]{Pike1998,Curdt2011,Mark2015}. This spatial pattern of Doppler velocity/Dopplergram reveals the rotating nature of the jet plasma column. In addition, the variations of Doppler velocity across the jet is also a signature of rotating motion of jet plasma column (e.g., \citealt{Young2014,Pariat2010}).\\
\begin{figure*}
\centering
\mbox{
\includegraphics[trim = 2.0cm 1.0cm 4.5cm 0.0cm, scale=1.0]{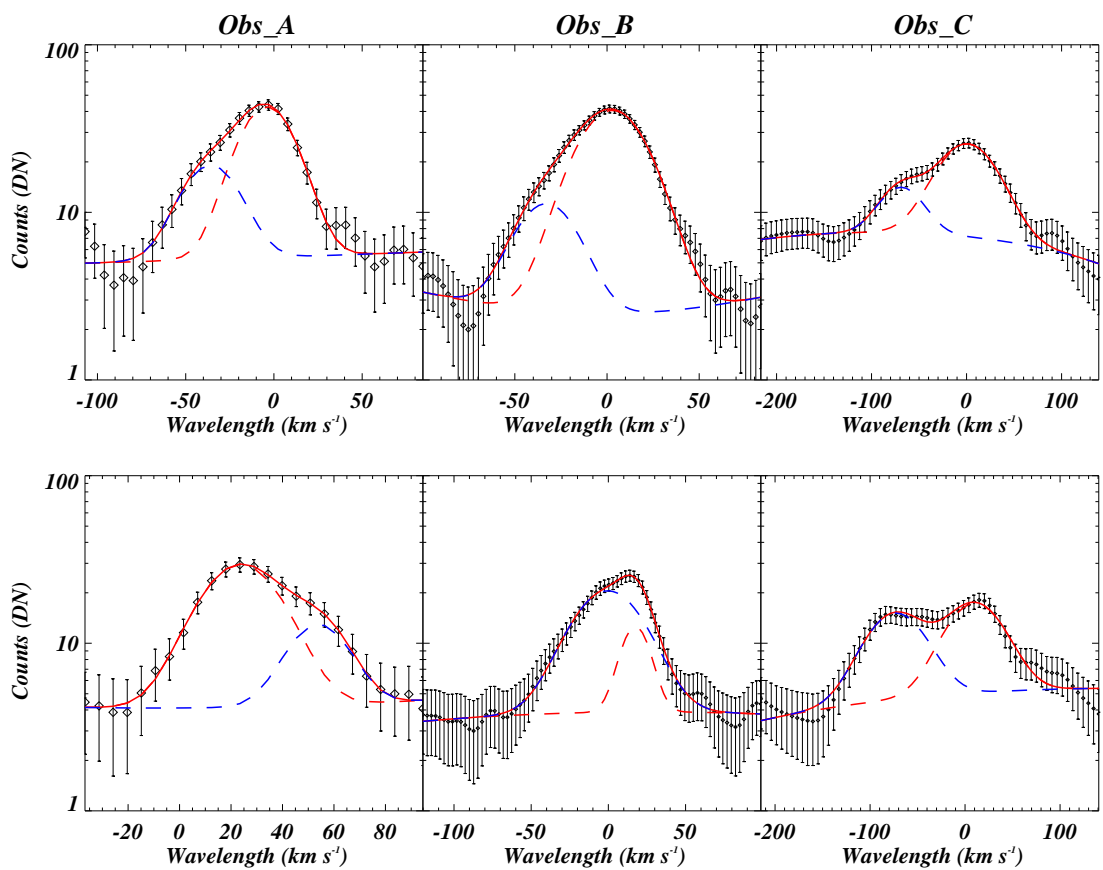}
}
\caption{\small The figure shows some sample spectral profiles and their fitting from Obs$\_$A (left-column), Obs$\_$B (middle column) and Obs$\_$ C (right-column). The black diamonds show the observed profiles while the solid red line represents the corresponding fitting. In addition, red dashed line shows the main Gaussian while blue dashed line shows the secondary Gaussian. The double Gaussian fitting leads to the much reliable fitting on the observed profiles.}
\label{fig:sample_spectra}
\end{figure*}
We utilize optically thin TR line (i.e., Si~{\sc iv} 1393.75~{\AA}) to understand the Doppler velocity pattern for these network jets. We have selected Si~{\sc iv} 1393.75~{\AA} spectral profiles across these network jets for each jet. After selecting these spectral profiles, we have averaged these spectral profiles using running average of 2 consecutive spectral profiles to increase the signal-to-noise ratio. We have found that the spectral profiles are significantly asymmetric within the vicinity of the network jets. The spectral profile appears as double peak profile (bottom-most panel; Fig.~\ref{fig:sample_spectra}) in some network jets. So, we have used double Gaussian (i.e., weak and main) fitting to characterize the observed line. Few sample profiles along with their double Gaussian fitting are shown in Fig.~\ref{fig:sample_spectra} from Obs$\_$A (left column), Obs$\_$B (middle column) and Obs$\_$C (right column). \cite{Young2014} have reported the occurrence of weak Gaussian (towards high velocity wing) along with the main Gaussian, which occurs in polar jets due to their very high speed. So, the secondary Gaussian basically contributes to the asymmetry of the line. The network jets are also very high speed plasma structure within the TR, which lead to the secondary Gaussian along with main peak of Si~{\sc iv} 1393.75~{\AA}. The high velocity component of this TR line is directly attributed from very high speed of network jets. Therefore, the secondary Gaussian of Si~{\sc iv} 1393.75~{\AA} line shows the true line-of-sight (LOS) Doppler velocity of these network jets. Interestingly, we have found that the double Gaussian fitting leads to very much reliable fitting of the observed profiles (cf, Fig.~\ref{fig:sample_spectra}).   
\begin{figure*}
\centering
\mbox{
\includegraphics[trim = 5.0cm 1.0cm 3.5cm 0.0cm, scale=0.45]{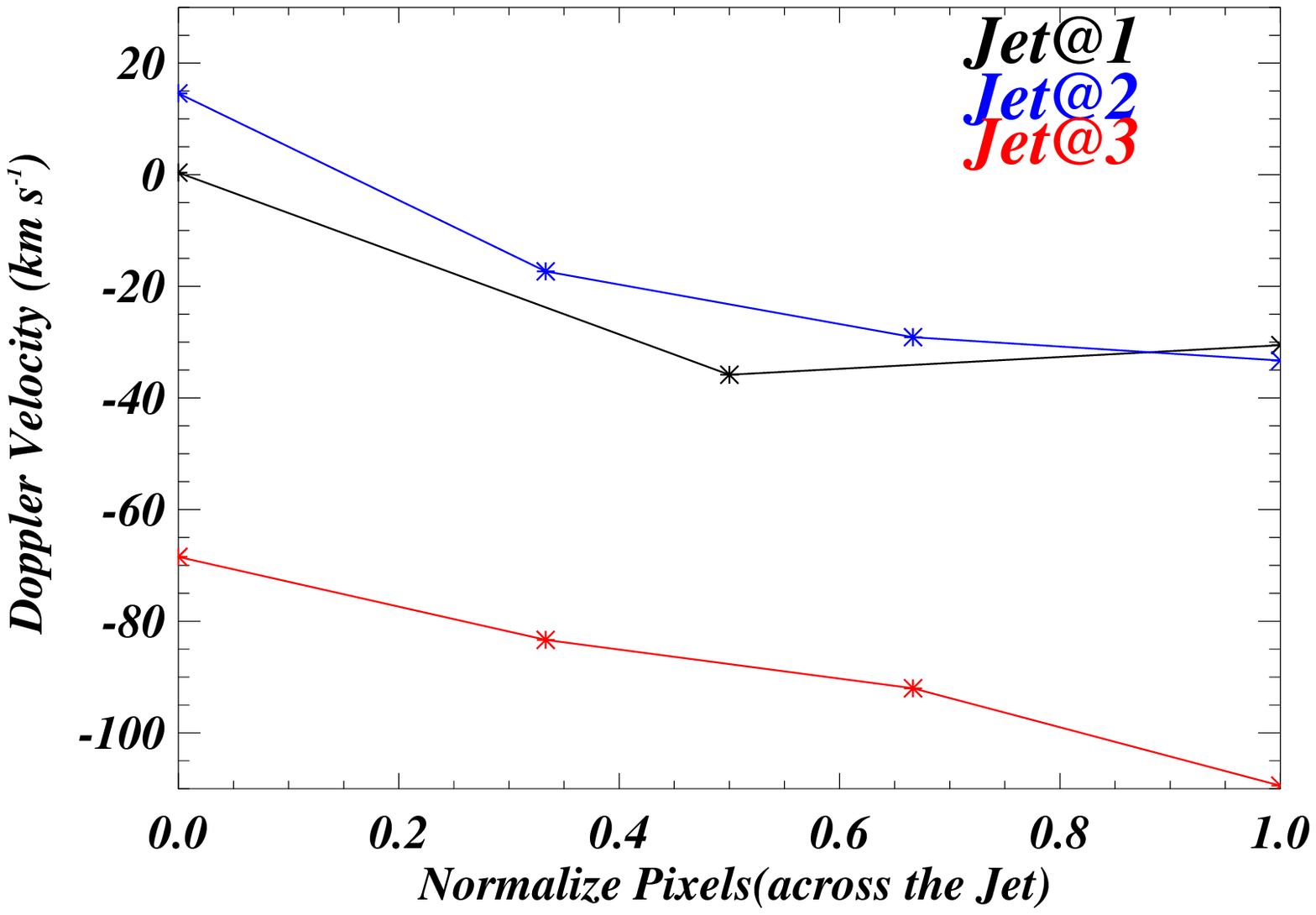}
\includegraphics[trim = -2.0cm 1.0cm 4.5cm 0.0cm, scale=0.45]{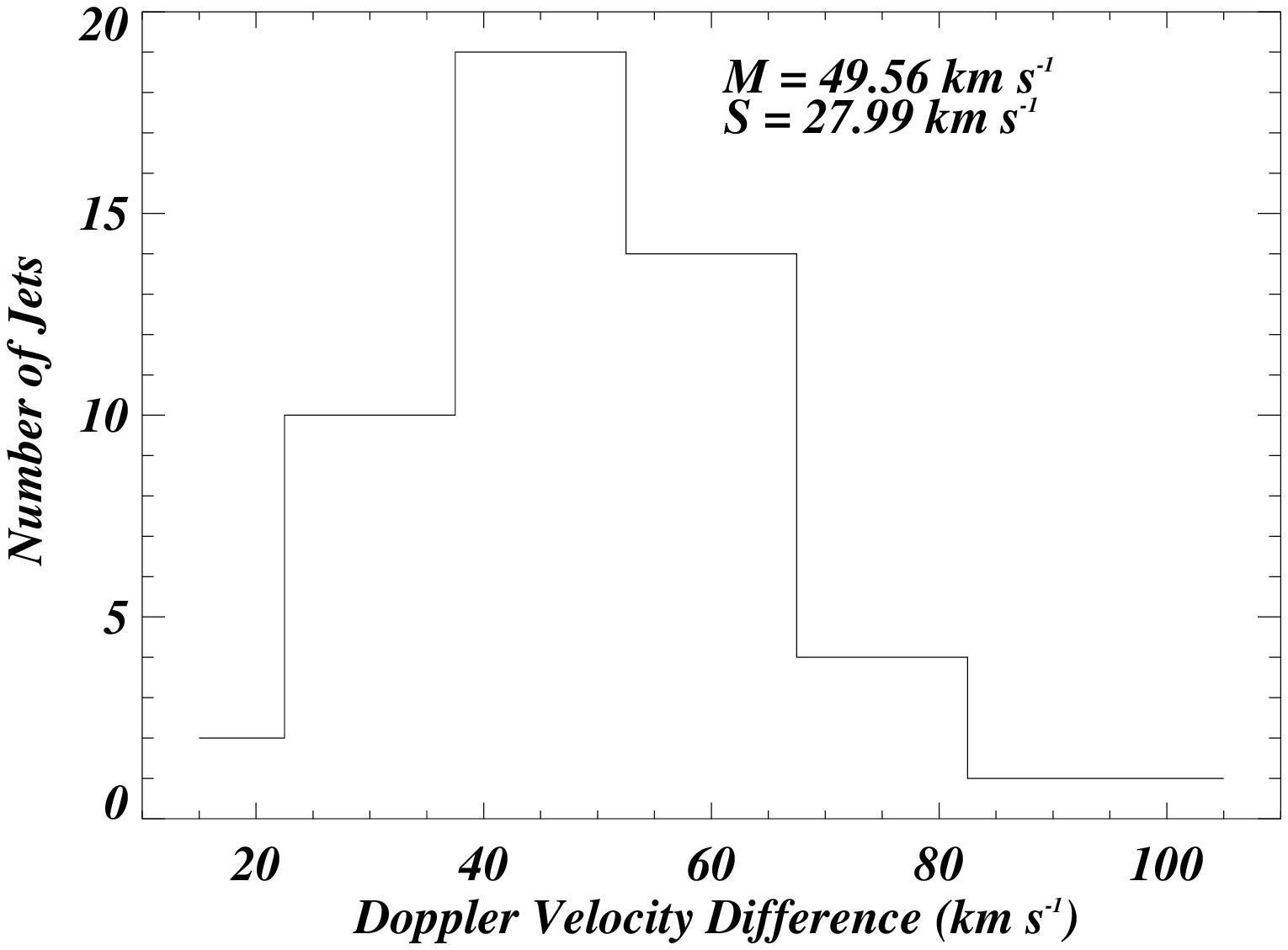}
}

\mbox{
}
\caption{\small The left-panel shows variation of Doppler velocity across the first jet (Obs$\_$A; black line), second jet (Obs$\_$B; blue line) and third jet (Obs$\_$C; red line). We have shown the distribution of $\Delta$V in the right-panel. The mean $\Delta$V is 49.56 km s$^{-1}$ with its standard deviation of 27.99 km s$^{-1}$.}
\label{fig:rot_mot}
\end{figure*}
In Fig.~\ref{fig:sample_spectra}, the black diamonds show the observed profile along with their errors. The over plotted solid red line represents the fitted profile. However, the red-dashed line shows the main Gaussian and blue-dashed line shows the secondary Gaussian. So, from all these displayed spectral profiles, we can see that the double Gaussian fitting leads to very much reliable fitting on these observed spectral profiles.\\
The variation of Doppler velocity across the jet is extremely important in the context of their rotational motion. In the left-panel of Fig.~\ref{fig:rot_mot}, the variations of Doppler velocity across the first jet (Obs$\_$A-jet@1; black solid line), second jet (Obs$\_$B-jet@2; blue line) and third jet (Obs$\_$C-jet@3; red line) are shown. First and second jets (jet@1 and jet@2) show that the red shift inverts into the blue shift from one edge of the jet to another edge, which is a typical signature of the rotational motion (e.g., \citealt{Pike1998,Curdt2011,Mark2015}). However, the jet@3 shows the increase in the Doppler velocity from one edge to its another edge. The increase or decrease of the Doppler velocity from one to another edge also signifies the presence of rotational motion within the jet body (\citealt{Young2014}). Using this procedure, we have investigated the Doppler velocity across these jets. It is found that in most of the network jets, the typical spatial pattern of Doppler velocity (i.e., blue shifts on one edge and red shifts on another edge) \textbf{emerge}. However, others have significant variations of the Doppler velocity from one edge of network jets to their another edge. Therefore, these results successfully predict the omnipresence of the rotational motion within these network jets.\\
To quantify the rational motion of these network jets, we took the difference of the Doppler velocity ($\Delta$ V) between edges of any particular jet. \cite{Young2014} have demonstrated that the $\Delta$V reflects the amount the rotation inherited in any particular network jet. We have estimated the $\Delta$V for each network jet to investigated their distribution. Finally, we have produced the histogram of $\Delta$V (right-panel; Fig.~\ref{fig:rot_mot}, which shows the mean value of $\Delta$V is 49.56 km s$^{-1}$ with their standard deviation of 27.99 km s$^{-1}$. However, the $\Delta$V can vary from 20.0 to 100.0 km s$^{-1}$. The angle between the jet's axis and observer (LOS direction) must be known to estimate the angular speed. However, the estimation of this angle (angle between jet axis and observer) is not possible from the used observational data. It should be noted that the angular speed is directly proportional to the $\Delta$V, therefore, it reflects the amount of rotational motion inherited in these network jets.  
\section{Discussion $\&$ Conclusions}\label{section:dis_con}
The high resolution imaging observations of TR from IRIS reveal the ubiquitous presence of network jets. We have used three different IRIS observations of QS, which are located near the disk center. On the basis of careful inspection, 51 network jets are identified from three QS observations and used for further analysis. These 51 network jets are very well resolved and not affected by the dynamics of other jets. The study is focused on the rotating motion of network jets along with the estimation of their other properties (e.g., speed, height and life-time). The mean speed, as predicted by statistical distributions of the speed, is 140.16 km s$^{-1}$ with standard deviation of 39.41 km s$^{-1}$. The mean speed of network jets is very similar as reported in the previous works (e.g., \citealt{Tian2014,Narang2016}). However, in case of their life-time, we found almost double mean lifetime (105.49 s) than the previously reported mean lifetime of the network jets (49.6 s; \citealt{Tian2014}). As we state earlier that we took only those network jet, which are very well resolved in the space as well as in the time, this criterion excludes short lifetime network jets. Therefore, our statistical distribution of the lifetime predicts higher mean lifetime. The mean length of the network jets is 3.16 Mm with its standard deviation of 1.18 Mm. In case of CH network jets, \cite{Tian2014} have reported that most of the network jets have length from 4.0 to 10.0 Mm. However, the mean length for QS network jets are smaller (3.53 Mm; \cite{Narang2016}). So, the mean length for QS network jets from the present work is well in agreement with \cite{Narang2016}. In addition, the apparent speed and length of these network jets are positively correlated, which is very similar as already reported in previous works (\citealt{Narang2016}). Finally, we can say that these networks jets are very dynamic features of the solar TR as revealed by their estimated properties.\\
The spectral profiles from TR are investigated extensively using space based observations. There are some noticeable features in spectral profiles of the TR, e.g., two distinct satellites to the blue and red indicating bi-directional flows during explosive events (\citealt{Dere1989}), enhanced emission in the wings above the networks (\citealt{Peter2000}). In addition, \cite{Peter2000} have demonstrated that the double Gaussian fitting yields the reliable fitting on TR spectral profiles and the secondary component is much more informative regarding the ongoing physical process. Recently, \citealt{Peter2010} have reported the asymmetry in coronal extreme ultraviolet lines in the vicinity of an active-region. The Si~{\sc iv} 1393.75~{\AA} spectral profiles are significantly asymmetric within the observed network jets. The presence of high plasma flow at any particular jet produces the secondary component along with the  main component of the line-profile, which leads to the asymmetry of the spectral lines (\citealt{Peter2010}). Therefore, the secondary component of the Si~{\sc iv} 1393.75~{\AA} line as observed in 51 jets in the present work, is most likely the result of high speed plasma flows (i.e., network jets). The LOS Doppler velocity of the secondary component represents the real LOS Doppler velocity from these networks jets. The occurrence of secondary component within the network jets is reported first time in our present work. Our study shows that most of the network jets have opposite Doppler shifts on their edges, which is a typical signature of rotating motion of the jet plasma column (e.g., \citealt{Pike1998,Curdt2011,Mark2015}). In addition, higher LOS Doppler velocity on one side than the Doppler velocity of other side also predicts the rotational motion (e.g., \citealt{Pariat2010,Young2014}) in these observed jets. We have found that some network jets show
this pattern (i.e., higher LOS Doppler velocity on one side than the other side Doppler velocity), which also justify their rotational motion. So, it is clear that all the observed network jets have rotating motion. The statistical analysis predicts the mean rotational motion (i.e., $\Delta$V) is 49.56 km s$^{-1}$ with its standard deviation 28.78 km s$^{-1}$. In case of polar jet, \cite{Young2014} reported that $\Delta$V $\approx$ 60.0 km s$^{1}$ (similar to the network jets) with its width 4.5 arcsec. However, in the present analysis, the statistical analysis of widths shows that the mean width is 0.62 arcsec, which is almost seven times lower than the width of polar jet. Qualitatively we can say that the angular speed of these network jets are higher than the usual solar jets (e.g., \citealt{Shen2011,Chen2012,Young2014}). Therefore, these network jets consist more helical magnetic fields than the other solar jets.\\
In addition, the observed properties of these network jets also help us to speculate on their triggering mechanism. The magnetic reconnection between the twisted magnetic field and pre-existing magnetic field may trigger the rotating motion (e.g., \citealt{Par2009, Fang2014,Pariat2016}). In a numerical model (\cite{Fang2014}), the magnetic reconnection between twisted magnetic fields and pre-existing magnetic field sheds the twist on newly reconnected field lines, and plasma flows along these twisted magnetic field lines. However, an another numerical model(\cite{Par2009})is based on photospheric motion, which can inherit the twist onto the magnetic field lines and finally produce the helical jet after the magnetic reconnection due to loosing of equilibrium of the flux-system. The present analysis predicts the ubiquitous presence of twist within these network jets. We have also found that the network jets have recurrent nature (i.e., many jets are triggered from the same location), which may be the result of oscillatory magnetic reconnection as proposed by \cite{Murray2009} and \cite{McL2012}. In addition, \cite{Goodman2014} have reported that Lorentz-force (magnetic acceleration) driven jets can have speed from 66-397 km s$^{-1}$. However, the pressure driven jets can also achieve the maximum speed of $\approx$ 60 km s$^{-1}$ (e.g.,\citealt{Mart2011}). \textbf{We tend to believe that the pressure driven jet may be able to account for the speed, but not the rotational motion}. Therefore, the rotating motion, recurrent nature and apparent high speed of the observed network jets (140.16 km s$^{-1}$) suggest that the magnetic reconnection  is the most-likely triggering mechanism in the present case. The similar aspect of the triggering mechanisms (i.e., recurrent nature and high apparent speed) of these network jets have also been reported in some previous studies, which confirm the occurrence of magnetic reconnection in support of the formation of network jets (e.g., \citealt{Tian2014,Narang2016}). However, very few network jets with the lower velocities (i.e., less than 60.0 km s$^{-1}$) may be formed due to gas pressure acceleration (\citealt{Shibata1982}).\\
We conclude that spectral analysis predicts the omnipresence of rotational motion in the network jets, which is reported first time for this class of jet-like structures. The helicity (amount of rotation) is high in the observed network jets compared to the usual other solar jets. In addition, the magnetic reconnection/acceleration is the most-likely cause behind the formation of these network jets. 

\begin{acknowledgements}
PK’s and KM's work was done in the framework of the project from the National Science Centre, Poland (NCN), Grant No. 2014/15/B/ST9/00106. IRIS is a NASA small explorer mission developed and operated by LMSAL with mission operations executed at NASA Ames Research center and major contributions to downlink communications funded by ESA and the Norwegian Space Centre. We also acknowledge the use of SDO/AIA observations for this study. The data provided are courtesy of NASA/SDO, LMSAL, and the AIA, EVE, and HMI science teams. AKS and BND acknowledge the RESPOND-ISRO project, while AKS acknowledges the SERB-DST young scientist project grant.
\end{acknowledgements}
\bibliographystyle{aa}

\bibliography{reference}
\end{document}